\documentclass[reqno]{amsart}
\usepackage{amsmath}
\usepackage{amssymb}
\usepackage{paralist}
\usepackage[colorlinks=true]{hyperref}
\hypersetup{urlcolor=blue, citecolor=red}

\theoremstyle{definition}

\numberwithin{equation}{section}
\numberwithin{theorem}{section}

\newcommand{\mc}[1]{{\mathcal #1}}

\newcommand{\bb}[1]{{\mathbb #1}}

\newcommand{\rme}{\mathrm{e}}

\newcommand{\rmd}{\mathrm{d}}
\newcommand{\eps}{\varepsilon}

\date{\today}
\title[From particle systems to the BGK equation]{From particle systems to the BGK equation}

\author[P.\ Butt\`a]{Paolo Butt\`a}
\address{Paolo Butt\`a\hfill\break \indent 
Dipartimento di Matematica, 
Sapienza Universit\`a di Roma,
\hfill\break \indent
P.le Aldo Moro 5, 00185 Roma, Italy}
\email{butta@mat.uniroma1.it}
\author[M.\ Pulvirenti]{Mario Pulvirenti}
\address{Mario Pulvirenti\hfill\break \indent
Dipartimento di Matematica, 
Sapienza Universit\`a di Roma,
\hfill\break \indent
P.le Aldo Moro 5, 00185 Roma, Italy}
\email{pulviren@mat.uniroma1.it}
\author[S.\ Simonella]{Sergio Simonella}
\address{Sergio Simonella\hfill\break \indent
ENS de Lyon, UMPA UMR 5669 CNRS
\hfill\break\indent
46 all\'{e}e d'Italie, 69364 Lyon Cedex 07, France
}
\email{sergio.simonella@ens-lyon.fr}

\begin{document}

\begin{abstract}
In \cite{BGK} the authors introduced a kinetic equation (the BGK equation), effective in physical situations where the Knudsen number is small compared to the scales where Boltzmann's equation can be applied, but not enough for using hydrodynamic equations. In this paper, we consider the stochastic particle system (inhomogeneous Kac model) underlying Bird's direct simulation Monte Carlo method (DSMC), with tuning of the scaled variables yielding kinetic and/or hydrodynamic descriptions. Although the BGK equation cannot be obtained from pure scaling, it does follow from a simple modification of the dynamics. This is proposed as a mathematical interpretation of some arguments in \cite{BGK}, complementing previous results in \cite{BHP,BP}.
\end{abstract}



\maketitle


\section{Introduction}
\label{sec:1}

In 1953 Bhatnagar, Gross and Krook \cite {BGK} proposed a new kinetic equation giving a tool of analysis, more efficient than the Boltzmann equation when the Knudsen number is small compared to the macroscopic scales, but not small enough to neglect the typical kinetic behaviour in favour of the hydrodynamic description given by the Euler equations. Hydrodynamics deals with the slow evolution of fields parametrizing the local equilibrium, which is typically established in a (much shorter) kinetic scale of time. Maintaining the description given by the Boltzmann equation, as far as practical questions are in focus, we are led to perform complex dynamical calculations (e.g., numerically) to obtain precise information on such local equilibria. One is tempted to simplify this task, replacing the two-body collision by an instantaneous thermalization on a local Maxwellian, constructed with the empirical parameters given by the dynamics itself. The equation for the one-particle distribution function $f=f(x,v,t)$ proposed in \cite{BGK} reads (neglecting mean field effects such as electric fields and external forces)
\begin{equation}
\label{BGK}
(\partial_t f + v\cdot \nabla_x f) (x,v,t) = \varrho(\varrho M_f - f) (x,v,t)\,,
\end{equation}
where
\begin{equation}
\label{Max}
M_f(x,v,t) = \frac 1 {(2 \pi T(x,t))^{3/2} } \exp\left(-\frac {|v-u(x,t)|^2}{2T(x,t)}\right)\,,
\end{equation}
and
\begin{equation}
\label{rtuf}
\begin{split}
& \varrho (x,t) = \int\! \rmd v\, f(x,v,t)\,, \quad \varrho \, u (x,t) = \int\! \rmd v\, f(x,v,t) v\,, \\ & \varrho(|u|^2 +3T)(x,t)  = \int\! \rmd v\, f(x,v,t) |v|^2 \,.
\end{split}
\end{equation}
Here, we fix the space dimension $d=3$, $(x,v)$ denotes position and velocity of a typical particle, and $t$ is the time. The Maxwellian $M_f$ has hydrodynamic parameters (density, mean velocity and temperature) obtained from local averages of $f$ itself. 

It turns out that \eqref{BGK}  has the same qualitative hydrodynamic behaviour of the Boltzmann equation, although the details of the interaction do not appear anymore in the evolution. In practice, \eqref{BGK} is not used to give a better approximation to the hydrodynamics, but, with respect to the Boltzmann equation, it is a simpler and more flexible tool to perform computations \cite{Ce88, So02}. 

We do not review here in any detail the very extensive literature (mathematical and applied) concerning BGK models. This includes numerical methods, hydrodynamic limits (see \cite{S-Ra03}, or \cite{Bi19} for a more recent contribution), analysis of  non-equilibrium steady states (as in \cite{Uk92,CELMM19,EM21}), or applications to gas mixtures (e.g., \cite{AAP02,BBGSP18,BIP19}), to name a few topics only.

The scope of the present paper is to suggest a mathematical derivation of \eqref{BGK} in terms of a minimal modification of a  stochastic particle model, introduced in Section \ref{sec:2} (a spatially inhomogeneous Kac model), which is commonly used in kinetic theory for the justification of Monte Carlo numerical schemes (such as the DSMC) in suitable scaling limits.

In two recent papers \cite{BHP,BP} the convergence of ad hoc stochastic particle systems to the solutions of the BGK equation \eqref{BGK} has been proved. Such particle systems are very different from the microscopic dynamics introduced below. Yet another, two-species particle system yielding the linear and homogeneous BGK equation rigorously has been recently studied in \cite{MW}.

The BGK equation is frequently used in the physics community as an efficient tool of computation, while the mathematical community considers it mostly as a toy model. We believe that the BGK equation has interesting aspects from the point of view of mathematical physics, which would deserve further investigation. We hope our discussion to be a step in this direction. 

The present analysis is purely formal. 
A rigorous approach would require considerable additional work starting, first of all, from constructive existence and uniqueness theorems for the solution of Eq.~\eqref{BGK}. At the moment, such results are available only when the first $\varrho$ on the right-hand side of \eqref{BGK} is replaced by a constant \cite{Pe89,PP} (although they can be extended to the case when $\varrho$ is replaced by a bounded function $\lambda (\varrho)>0$, and this is a reasonable  physical assumption).

\section{Basic particle systems and their kinetic li\-mits}
\label{sec:2}

Let  $\bb T_\ell^3$ be the $3$-dimensional torus of side $\ell$. We consider a system of $N$ identical  particles in $\bb T_\ell^3$ and denote by $Z_N = (X_N,V_N )$ a configuration of the system, where $X_N = (x_1,\ldots,x_N) \in (\bb T_\ell^3)^N $ and $V_N = (v_1,\ldots,v_N) \in (\bb R^3)^N$  are the positions and velocities of the particles, respectively. We shall also use the notation $Z_N = (z_1,\ldots,z_N)$  with $z_j = (x_j,v_j) $. The particles move according to the following stochastic dynamics. They are moving freely until a random Poisson time of intensity scaling as $\frac{N(N-1)}{2}$, when a pair of them is extracted with an equal probability scaling as $\frac{2}{N(N-1)}$. If the particles of such pair are at a distance less than one, they perform an elastic collision with a random impact parameter $\omega$. Otherwise, nothing happens. More precisely, if $\Phi=\Phi(Z_N)$ is a test function on the state space, the generator of the process reads, in microscopic variables,
\[
\begin{split}
\mc L_\mathrm{m} \Phi (X_N,V_N ) & = V_N \cdot \nabla_{X_N}  \Phi (X_N,V_N) + \sum_{i<j} \int\! \rmd\omega\, B(\omega;v_i-v_j) \\ & \quad \times  \varphi (|x_i-x_j|) \{ \Phi (X_N,V_N^{i,j} ) - \Phi(X_N,V_N)\}\,.
\end{split}
\]
Here $\varphi (r)$ is supported in $(0,1)$ and can be taken, for simplicity, as the characteristic function of such set; $V_N^{i,j}$ has the same components of $V_N$ but for $v_i$ and $v_j$, which are replaced by the outgoing velocities $v'_i$ and $v'_j$ of a collision law with incoming velocities $v_i$ and $v_j$ and impact parameter $\omega$,
\[
\begin{cases}
v'_i  = v_i -  \left((v_i-v_j)\cdot \omega \right) \omega\,,  \\
v'_j  = v_j +  \left((v_i-v_j)\cdot \omega \right) \omega\,.
\end{cases}
\]
Finally, $B>0$ is chosen as the cross-section of the Maxwell molecules with angular cutoff for which
\[
\int\! \rmd\omega\, B(\omega;V) = 1\,.
\]

Up to now we are arguing in terms of microscopic variables, in which the size $\ell$ of the configuration space $\bb T_\ell^3$ is very large. Introducing now the space-time scale parameter $\eps=\ell^{-1}>0$, we pass to macroscopic variables
\[
x \to \eps x\,, \quad  t \to \eps t \,,
\]
which belong to the unit torus  $\bb T_1^3 =: \bb T ^3$. In the low-density regime, one assumes
\begin{equation}
\label{scal}
\eps^2 N = 1\,.
\end{equation}
In the macroscopic variables, the generator takes the form
\[
\mc L_\mathrm{m} \Phi (X_N,V_N )= V_N \cdot \nabla_{X_N}  \Phi (X_N,V_N) + \mc L_\mathrm{int} \Phi (X_N,V_N )\,,
\]
where
\[
\begin{split}
\mc L_\mathrm{int} \Phi (X_N,V_N ) & = \eps^2 \sum_{i<j} \int\! \rmd\omega\, B(\omega;v_i-v_j) \varphi_\eps (|x_i-x_j|) \\ & \quad \times \{ \Phi (X_N,V_N^{i,j} )- \Phi(X_N,V_N)\}
\end{split}
\]
and
\[
\varphi_\eps(r) = \frac 1 {\eps^3} \varphi \left(\frac r\eps\right)
\]
is an approximation of the delta function.
The formal link with the Boltzmann equation is explained next.

Consider a symmetric probability distribution $W^N(Z_N,t)$ solution to the master equation (forward Kolmogorov equation),
\begin{equation}
\label{master}
(\partial_t +V_N \cdot \nabla_{X_N}) W^N(Z_N,t) =  \mc L_\mathrm{int} W^N (Z_N,t )\,.
\end{equation}
From this we can obtain a hierarchy of equations for the marginals associated to $W^N$. In particular, denoting by $f^N_1$ and $f^N_2$ the one and two particle marginals, the first hierarchical equation is
\begin{align}
\label{ger1}
& \partial_t f^N_1(x_1,v_1,t) + v_1 \cdot \nabla_{x_1} f^N_1(x_1,v_1,t) \nonumber \\ & \qquad = \eps^2 (N-1) \int\! \rmd\omega\! \int\! \rmd x_2\! \int\! \rmd v_2\,  B(\omega;v_1-v_2) \varphi_\eps (x_1-x_2) \nonumber \\ & \qquad \quad \times  \{f_2^N (x_1, v_1',x_2, v_2',t) - f_2^N (x_1, v_1,x_2, v_2,t)\}\,.
\end{align} 
If $W^N$ is initially chaotic, namely $W^N(Z_N, 0) = f_0^{\otimes N}(Z_N)$, assuming that in the limit $N\to \infty$ propagation of chaos occurs at any positive time and taking $\eps$ as in \eqref{scal}, from \eqref{ger1} we formally obtain the Boltzmann equation.
A mathematical rigorization of this argument is not obvious at all\footnote{One can apply the method of Lanford for mechanical systems \cite{L} to obtain a short time validity result working in $L^{\infty}(\bb T^3 \times \bb R^3)$ (and assuming fast velocity decay). Unfortunately, we cannot approach the problem  in $L^{1}(\bb T^3 \times \bb R^3)$ because, due to the presence of $\varphi_\eps$, the collision operator has an $L^1$-norm diverging with $\eps$.}.

In spite of the presence of the factor $\eps^2 = 1/N$ in the interaction operator, what we are dealing with is far from a mean-field model. Actually, the model is rather intractable both at the mathematical and at the practical level, at least at the scales of time of interest in the applications. It is indeed close to the more fundamental, Hamiltonian system of deterministic particles following the Newton's law.

The BGK equation cannot follow directly from the previous model, not even modifying the scaling relation \eqref{scal}. In fact, when
\begin{equation}
\label{scalh}
\eps^{\alpha}N = 1 \quad \text{for }\;\; \alpha \in (2, 3]
\end{equation}
we obtain hydrodynamic equations for the slow time evolution of the fields which parametrize the local equilibria.

Notice that, in the scaling \eqref{scal}, the average number of particles falling in the ball $B_\eps(x_1)$ of radius $\eps$ around $x_1$ is $o(1)$, so that it is difficult to figure out the instantaneous thermalization which is present in the BGK model. Therefore, a natural proposal is a mean-field particle model in which either $\varphi_\eps = \varphi$  is independent of $\eps$ or it approximates the delta function much more gently. We will do so by introducing a partition of the torus in square cubes, exactly in the spirit of classical numerical codes \cite{Bi94}.

\section{The mean-field stochastic particle system}
\label{sec:3}

Let $\{\Delta\}$ be  a partition of $\bb T^3$ in cubic cells $\Delta$ with equal volume $|\Delta|$. Consider a system of $N$ particles evolving freely in $\bb T^3$ up to an exponential time of suitable intensity. At such time, a pair of particles is extracted randomly. If they fall in the same cell  $\Delta$, they may perform a collision as in the basic system of Section \ref{sec:2}. Otherwise, nothing happens. 

As before, $Z_N=(X_N,V_N)=(z_1, \ldots, z_N)$ denotes a configuration of the system, being $z_i=(x_i,v_i)$ position and velocity of the $i$-th particle. The generator of this process reads ($\Phi = \Phi(Z_N)$ a test function)
\begin{align}
\label{SP}
\mc L \Phi (Z_N) & = V_N \cdot \nabla_{X_N} \Phi (Z_N) +
\frac 1{N |\Delta|} \sum_{i<j} \int\! \rmd\omega\, B(\omega;v_i-v_j) \chi_{i,j} \nonumber \\ & \quad \times \{\Phi (X_N, V_N^{i,j})- \Phi(X_N,V_N)\}\,,
\end{align}
where $\chi_{i,j} =1$ if $i$ and $j$ belong to the same cell and $0$ otherwise. As before, we denote by $W^N(Z_N,t)$ a symmetric probability distribution solution to the associated master equation,
\begin{align}
\label{master1}
& (\partial_t +V_N \cdot \nabla_{X_N}) W^N(Z_N,t) \nonumber \\ & = \frac 1{N |\Delta|} \sum_{i<j} \int\! \rmd\omega\, B(\omega;v_i-v_j) \chi_{i,j} \{W^N(X_N, V_N^{i,j})- W^N(X_N,V_N)\}\,.
\end{align}
There exist several variants of such spatially inhomogeneous, mean-field particle models with collisions. For instance, Cercignani's model of soft spheres \cite{Ce83,LP90} in which, at variance with the above proposal, the impact vector $\omega$ is not random; we refer to \cite{PPS19} for an account of related mathematical results.

This process yields formally the Boltzmann equation in the combined limit $N\to \infty$ and $|\Delta|\to 0$. Indeed, let  $W^N(Z_N,t)$ be a symmetric probability distribution solution to the master equation \eqref{master1}. If $f^N_1$ and $f^N_2$ are the one and two particle marginals, for any test function $\varphi=\varphi(z)$, 
\begin{align}
\frac{\rmd}{\rmd t} \int\! \rmd z_1\, f^N_1 \varphi & = \int\! \rmd z_1\, f^N_1 v_1 \cdot \nabla_x \varphi + \frac {N-1}{N|\Delta|} \int\! \rmd z_1\,\rmd z_2\! \int\! \rmd\omega\,  B(\omega;v_1-v_2) \nonumber \\ & \quad \times \chi_{1,2} f^N_2 (z_1,z_2)   \{ \varphi (x_1, v_1')- \varphi (x_1, v_1)\}\,.
\end{align}
Therefore, under the assumption of propagation of chaos, letting first $N \to \infty$ and then $|\Delta| \to 0$ we recover the Boltzmann equation in the weak form (assuming the existence of a global solution and its stability with respect to a regularization via a cell partition).

\subsection{BGK equation}
\label{sec:3.1}

To derive, at least formally, the BGK model, we introduce a modification of the stochastic process \eqref{SP} in which, inspired from the original paper \cite{BGK}, we reinforce the interaction leaving finite the mean-free path. To do this, we introduce a time $\tau$, which will eventually converge to $0$, and prescribe the dynamics in each time interval $[2n\tau, 2(n+1)\tau]$, $n\in \bb N$, according to the following rules. All the particles move freely in the time interval $[2n\tau,(2n+1)\tau]$, while, during the time interval $[(2n+1)\tau, (2n+2)\tau]$, the particles contained in each cell $\Delta$ evolve according to the homogeneous Kac dynamics with probability $\tau N_\Delta/N$ and nothing happens with probability $1- \tau N_\Delta/N$, where $N_\Delta$ denotes the number of such particles. This allows to preserve the mean free path finite, being $\tau$ properly small. Moreover, we increase the number of collisions introducing a time-scale parameter $\eps$ in the Kac dynamics. 

The solution $W^N(t)$ to the corresponding master equation (hereafter, we will often omit the explicit dependence of $W^N$ on the variables $Z_N$) is thus given by a product formula,
\[
W^N(n\tau) = (S_0(\tau)K(\tau))^n W^N(0)\,,
\]
where $S_0$ is the free stream operator and
\[
K(\tau) =  \prod_{\Delta} \left[\frac{\tau N_\Delta}N S^\Delta(\tau) + \left( 1 - \frac{\tau N_\Delta}N \right) \right],
\]
with
\[
S^\Delta(\tau) = \exp\left(\frac{\tau}{\eps} \mc L_\mathrm{int}^\Delta \right)
\]
and ($G = G(V_N)$ a test function)
\[
\mc L_\mathrm{int}^\Delta G (V_N) = \frac 1{N |\Delta|} \sum_{i<j} \int\! \rmd\omega\, B(\omega;v_i-v_j) \chi^\Delta_{i,j} \{G(V_N^{i,j}) - G(V_N)\}\,.
\]
Above, $\chi^\Delta_{i,j}=1$ iff $x_i,x_j \in \Delta$, and $\chi^\Delta_{i,j}=0$ otherwise. Moreover, we assume $\eps \ll \tau \ll 1$.

The product formula is easily rewritten as a discrete time Duhamel formula with respect to the linear evolution $S_0$, 
\begin{align*}
W^N(n \tau) & = S_0(\tau) (K(\tau)-1) W^N ((n-1)\tau)  + S_0(\tau) W^N ((n-1)\tau) \\ & = \cdots \\ & =S_0(n\tau) W^N(0) + \sum_{k=1}^n S_0(k\tau) ( K(\tau) -1) W^N [ (n-k) \tau]\,.
\end{align*}
We next observe that 
\[
K(\tau)- 1 = \tau \sum_{\Delta} \frac{N_\Delta}N (S^\Delta(\tau) -1) + O(\tau^2)\,,
\]
whence, for small $\tau$,
\begin{equation}
\label{Duh}
W^N( n\tau) \approx S_0(n\tau) W^N(0) + \sum_{k=1}^n \tau S_0(k\tau) \sum_\Delta \frac{N_\Delta}N (S^\Delta(\tau) -1) W^N ((n-k) \tau)\,.
\end{equation}

Now, let $f^N_1$ be the one-particle marginal,
\[
f^N_1(t) = f^N_1(z_1, t) = \int\! \rmd Z_{1,N}\, W^N(Z_N, t)\,,
\]
being $\rmd Z_{1,N} = \rmd z_2 \cdots \rmd z_N$. Integrating both sides of \eqref{Duh} with respect to $\rmd Z_{1,N}$ and then changing variables $X_{1,N} \to X_{1,N} +  k\tau V_{1,N}$ we get
\begin{equation}
\label{DuWf}
f^N_1(n\tau) = s_0(n\tau)f^N_1(0) + \sum_{k=1}^n \tau s_0(k\tau)  QW^N ((n-k)\tau)\,,
\end{equation}
where $s_0$ is the one-particle free stream operator and 
\[
QW^N(t) = QW^N(z_1,t) = \int\! \rmd Z_{1,N}  \sum_\Delta  \frac{N_\Delta}N (S^\Delta (\tau) -1) W^N(Z_N,t)\,.
\]

We next write,
\[
\begin{split}
QW^N(z_1,t) & =  \int\! \rmd X_{1,N}\, R^N_t(X_N) \sum_\Delta  \frac{N_\Delta}N \\ & \quad \times \int\! \rmd V_{1,N}\, (S^\Delta(\tau)-1) \Pi^N_t(V_N|X_N)\,,
\end{split}
\]
with $R^N_t(X_N) = \int\! \rmd V_N\, W^N(Z_N,t)$ the spatial density and $\Pi^N_t(V_N|X_N)$ the distribution in velocity conditioned to $X_N$ (which, for the moment, plays the role of a parameter). Denoting by $V_N^A$ the velocity variables of the particles in $A\subset \bb T^3$, we set (with an abuse of notation)
\[
\Pi^N_t(V_N^{\Delta^c}| X_N) = \int\! \rmd V_{N}^\Delta\, \Pi^N_t(V_N | X_N)
\]
and let $\Pi^N_t (V_N^\Delta | X_N, V_N^{\Delta^c})$ be the distribution $\Pi^N_t(V_N|X_N)$ conditioned to $V_N^{\Delta^c}$, so that
\[
\Pi^N_t(V_N^{\Delta^c} | X_N) \Pi^N_t (V_N^\Delta | X_N, V_N^{\Delta^c}) = \Pi^N_t(V_N|X_N)\,.
\]
If $\Delta_1$ is the cell containing $x_1$, then for any $\Delta \ne \Delta_1$ we have
\[
\begin{split}
& \int\! \rmd V_{1,N}\, (S^\Delta(\tau)-1)  \Pi^N_t(V_N | X_N)  =  \int\! \rmd V_{1,N}^{\Delta^c}\, \Pi^N_t(V_N^{\Delta^c} | X_N) \\ & \qquad \times \int\! \rmd V_N^\Delta (S^{\Delta}  (\tau)-1)  \Pi^N_t (V_N^\Delta | X_N, V_N^{\Delta^c}) = 0\,,
\end{split}
\] 
having used, in the last equality, that $\rmd V_N^\Delta$ is stationary under $S^\Delta (\tau)$. Hence,
\begin{align}
\label{QW}
QW^N(z_1,t) & =   \int\! \rmd X_{1,N}\, R^N_t(X_N) \frac{N_{\Delta_1}}N \int\! \rmd V_{N}^{\Delta^c_1}\, \Pi^N_t(V_N^{\Delta_1^c} | X_N) \nonumber \\ & \quad \times \int\! \rmd V_{1,N}^{\Delta_1} (S^{\Delta_1}(\tau)-1) \Pi^N_t (V_N^{\Delta_1} | X_N, V_N^{\Delta_1^c})\,.
\end{align}
Now, for $\eps\ll \tau$, the mixing property of the Kac model 
implies
\[
S^{\Delta_1} (\tau) \Pi^N_t (V_N^{\Delta_1} | X_N, V_N^{\Delta_1^c}) \approx \mu_{\mc E^N_{\Delta_1},\mc P^N_{\Delta_1} } (V_N^{\Delta_1})\,,
\]
where $\mu_{\mc E^N_{\Delta_1},\mc P^N_{\Delta_1}}$ is the microcanonical measure associated to the empirical energy and momentum in $\Delta_1$,
\[
\mc E^N_{\Delta_1} = \frac 1 {2N_{\Delta_1}} \sum_{j:\, x_j \in \Delta_1} v_j^2\,, \quad \mc P^N_{\Delta_1} = \frac 1 {N_{\Delta_1}} \sum_{j:\, x_j \in \Delta_1} v_j\,.
\]
On the other hand, letting $\pi^N_{\Delta_1} = N_{\Delta_1}/N$ be the empirical density in $\Delta_1$, we expect that, with large (i.e., converging to one) $R^N_t(X_N)$-probability when increasing $N$,
\[
\pi^N_{\Delta_1} \approx  \varrho^N_{\Delta_1}(t) := \int_{\Delta_1}\!\rmd x\, \varrho^N_1(x,t)
\]
(where $\varrho^N_1(x,t) := \int\!\rmd v \, f^N_1(x,v,t)$) and that the $v_j$'s are asymptotically independent. Therefore, by the law of large numbers, again with large $R^N_t(X_N)$-probability when increasing $N$, we expect also
\[
\begin{split}
\mc E^N_{\Delta_1} & \approx E^N_{\Delta_1}(t) := \frac 1{\varrho^N_{\Delta_1}(t)} \int _{\Delta_1}\! \rmd x\! \int\! \rmd v\, f^N_1(x,v,t) \frac {v^2}2\,, \\ \mc P^N_{\Delta_1} & \approx P^N_{\Delta_1}(t) := \frac 1{\varrho^N_{\Delta_1}(t)}  \int _{\Delta_1}\! \rmd x\! \int\! \rmd v\, f^N_1(x,v,t) v\,.
\end{split}
\]
Since the functions $\varrho^N_{\Delta_1}$, $E^N_{\Delta_1}$, and  $P^N_{\Delta_1}$ are non random, inserting the above approximations in \eqref{QW} and using the obvious identities
\[
\begin{split}
& \int\! \rmd X_{1,N}\, R^N_t(X_N) \int\! \rmd V_{N}^{\Delta^c_1}\,\Pi^N_t(V_N^{\Delta_1^c} | X_N) =  \int\! \rmd X_{1,N}\, R^N_t(X_N) = \varrho^N_1(x_1,t) \,, \\ & \int\! \rmd Z_{1,N}\, R^N_t(X_N) \Pi^N_t(V_N^{\Delta_1^c} | X_N) \Pi^N_t (V_N^{\Delta_1} | X_N, V_N^{\Delta_1^c})  = f^N_1(z_1,t)\,,
\end{split}
\]
we obtain
\[
QW^N(z_1,t) \approx \varrho^N_{\Delta_1}(t) \bigg(\varrho^N_1(x_1,t) \int\! \rmd V_{1,N}^ {\Delta_1}\, \mu_{E^N_{\Delta_1}(t),P^N_{\Delta_1}(t)} (V_N^{\Delta_1}) - f^N_1(z_1,t) \bigg).
\]
We finally observe that by the equivalence of the ensembles, see Appendix \ref{sec:A}, the marginal distribution $ \int\! \rmd V_{1,N}^ {\Delta_1}\, \mu_{E^N_{\Delta_1},P^N_{\Delta_1}} (V_N^{\Delta_1})$ is close (since  $N_{\Delta_1} \approx \varrho^N_{\Delta_1}N$ is large) to the Maxwellian
\[
M_{P^N_{\Delta_1},T^N_{\Delta_1}} (v_1) = \frac 1{(2\pi T^N_{\Delta_1})^{3/2}} \exp\left( - \frac {|v_1-P^N_{\Delta_1})|^2} {2T^N_{\Delta_1}}\right),
\]
with $3T^N_{\Delta_1} = 2 E^N_{\Delta_1} - (P^N_{\Delta_1})^2$.  In conclusion,
\[
QW^N(z_1,t)  \approx \varrho^N_{\Delta_1}(t) \left(\varrho^N_1(x_1,t) M_{P^N_{\Delta_1}(t),T^N_{\Delta_1}(t)} (v_1) - f^N_1(z_1,t) \right).
\]

Inserting the last approximation for $QW^N$ in \eqref{DuWf} we get
\[
f^N_1(n\tau) \approx s_0(n\tau)f^N_1(0) + \sum_{k=1}^n \tau s_0(k\tau) [\varrho^N_{\Delta_1}  (\varrho^N_1 M_{P^N_{\Delta_1},T^N_{\Delta_1}} - f^N_1) ] ((n-k)\tau)\,.
\]
Taking the limits $\eps \to 0$, $N\to \infty$, and finally $\tau \to 0$, the above display implies that the (limit) one-particle marginal $f$ solves the integral equation,
\[
f(x,v,t) =  f_0(x-vt,v) + \int_0^t\!\rmd s\, \varrho_{\Delta_x} (\varrho M_{P_{\Delta_x},T_{\Delta_x}} - f) (x-vs,v,t-s)\,,
\]
where $\Delta_x$ is the cell containing $x$ and $\varrho_\Delta$, $P_\Delta$, $T_\Delta$ are defined as $\varrho^N_\Delta$, $P^N_\Delta$, $T^N_\Delta$ with $f^N_1$ replaced by $f$. 

Finally, taking also the limit $|\Delta|\to 0$, we recover the equation
\[
f(x,v,t) =  f_0(x-vt,v) + \int_0^t\!\rmd s\,(\varrho (\varrho M_f-f))(x-vs,v,t-s)\,,
\]
which is the mild formulation of Eq.~\eqref{BGK} via Duhamel formula.


\appendix
\section{Equivalence of ensembles}
\label{sec:A}

Let $\mu_N^{E,P}$ denote the uniform (i.e., microcanonical) distribution in
\[
\mc X_N^{E,P} := \bigg\{V_N\in (\bb R^3)^N \colon \sum_{j=1}^N |v_j|^2 = 2NE\,, \;\; \sum_{j=1}^N v_j = N P  \bigg\}.
\]
In \cite[Lemma 4.1]{CCL}, it is shown that if $N\ge 3$ and $V_N\in \mc X_N^{1/2,0}$ is distributed according to $\mu_N^{1/2,0}$ then each variable
\[
\tilde v_j = \sqrt{\frac{N}{N-1}}\, v_j\,, \quad j = 1, \ldots, N\,,
\]
is distributed in the unit ball of $\bb R^3$ with law
\[
\rmd \nu_N(v) = \frac{|S^{3N-7}|}{|S^{3N-4}|} \left(1-|v|^2\right)^{(3N-8)/2} \rmd v\,, 
\]
where $S^n$ denotes the unit $n$-sphere. From this we deduce that if $T$ is the temperature such that $2E-P^2 = 3T$, $N\ge 3$, and $V_N\in \mc X^{E,P}_N$ is distributed according to $\mu_N^{E,P}$, then each velocity $v_j$ is distributed in the ball of radius $\sqrt{3T(N-1)}$ and center $P$ with law 
\[
\rmd \nu_N^{E,P}(v) = \frac{1}{[3T(N-1)]^{3/2}} \frac{|S^{3N-7}|}{|S^{3N-4}|} \left(1- \frac{|v-P|^2}{3T(N-1)}\right)^{(3N-8)/2} \rmd v\,, 
\]
Recalling that 
\[
|S^n|=\frac{2\pi^{\frac{n+1}2}}{\Gamma\left(\frac{n+1}2\right)}\,,
\]
where $\Gamma(x)$ denotes the gamma function, and using the Stirling approximation 
\[
\Gamma(x) =  \sqrt{2\pi} \, x^{x-\frac 12} \,\rme^{-x} \left[ 1 + O\left(\frac 1x \right) \right], \quad x>0\,,
\]
we get
\[
\begin{split}
\frac{|S^{3N-7}|}{|S^{3N-4}|} & =  \frac{1}{(\pi\rme)^{3/2}}  \left(\frac{3N-4}2\right)^{\frac{3N-5}2} \left(\frac{3N-7}2\right)^{-\frac{3N-8}2} \left[ 1 + O\left(\frac 1N \right) \right] \\ & =  \frac{1}{(\pi\rme)^{3/2}} \exp\left(\frac{3N-5}{2} \log \frac{3N-4}{3N-7} + \frac 32 \log\frac{3N-7}2 \right) \\ & \quad \times \left[ 1 + O\left(\frac 1N \right) \right] =  \left(\frac{3N}{2\pi}\right)^{3/2}  \left[ 1 + O\left(\frac 1N \right) \right],
\end{split}
\]
so that
\[
\lim_{N\to \infty}  \frac{1}{[3T(N-1)]^{3/2}} \frac{|S^{3N-7}|}{|S^{3N-4}|} = \frac{1}{(2\pi T)^{3/2}}\,.
\]
On the other hand,
\[
\lim_{N\to \infty} \left(1- \frac{|v-p|^2}{3T(N-1)}\right)^{(3N-8)/2}  = \exp\left(-\frac{|v-P|^2}{2T} \right).
\]
Therefore, looking at $ \rmd \nu_N^{E,P}(v)$ as a probability on $\bb R^3$, its density 
\[
g_N^{E,P}(v) = \frac{1}{[3T(N-1)]^{3/2}} \frac{|S^{3N-7}|}{|S^{3N-4}|} \left(1- \frac{|v-P|^2}{3T(N-1)}\right)_+^{(3N-8)/2}
\]
converges pointwise to $M_{P,T}(v) = [2\pi T]^{-3/2} \exp\left(-\frac{|v-P|^2}{2T} \right)$ as $N\to \infty$. Moreover, there are $C_1,C_2>0$ such that $g_N^{E,P}(v) \le C_1 \exp \left(-C_2|v-P|^2\right)$ (this can be seen using, e.g., the inequality $1-r \le \rme^{-r}$ valid for any $r>0$), so that, by dominated convergence, we also have
\[
\lim_{N\to \infty} \int\!\rmd \nu_N^{E,P}(v)\, h(v) = \int\!\rmd v\, M_{P,T}(v)\,,
\] 
for any function $h\in L^1(\bb R^3; \rme^{C_2 v^2}\rmd v)$.


\end{document}